\documentclass[a4paper,11pt]{article}
\pdfoutput=1 
\usepackage{indentfirst}
\usepackage{jheppub} 
\usepackage[T1]{fontenc} 
\RequirePackage[displaymath]{lineno} 
\usepackage{soul}
\usepackage{epsfig}
\usepackage{graphicx}
\usepackage{dcolumn}
\usepackage{bm}
\usepackage{ltablex,booktabs}
\usepackage{overpic}
\usepackage{subfigure}
\usepackage{float}
\usepackage{color}
\usepackage{amsmath}
\usepackage{mathcomp}
\usepackage{mathrsfs}
\usepackage{multirow}
\usepackage{rotating}
\usepackage{amssymb}
\usepackage{gensymb}
\usepackage{amsmath}
\usepackage{tabularx}
\usepackage{epstopdf}
\usepackage[normalem]{ulem}
\newcommand{\dstoksk}{D_{s}^{-} \to K_{S}^{0}K^{-}}
\newcommand{\dstokkpi}{D^{-}_{s} \to K^{+}K^{-}\pi^{-}}

\newcommand{\dstopietapgam}{D^{-}_{s} \to \pi^{-}\eta^\prime_{\pi^{+}\pi^{-}\eta(\gamma \gamma)}}

\newcommand{\BESIIIorcid}[1]{\href{https://orcid.org/#1}{\hspace*{0.1em}\raisebox{-0.45ex}{\includegraphics[width=1em]{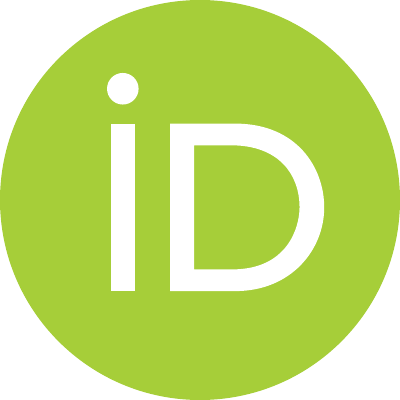}}}}

\title{Search for the baryonic decay $ D_{s}^{*+} \to \ p \bar{n} $}
\collaborationImg{\includegraphics[width=0.15\textwidth, angle=90]{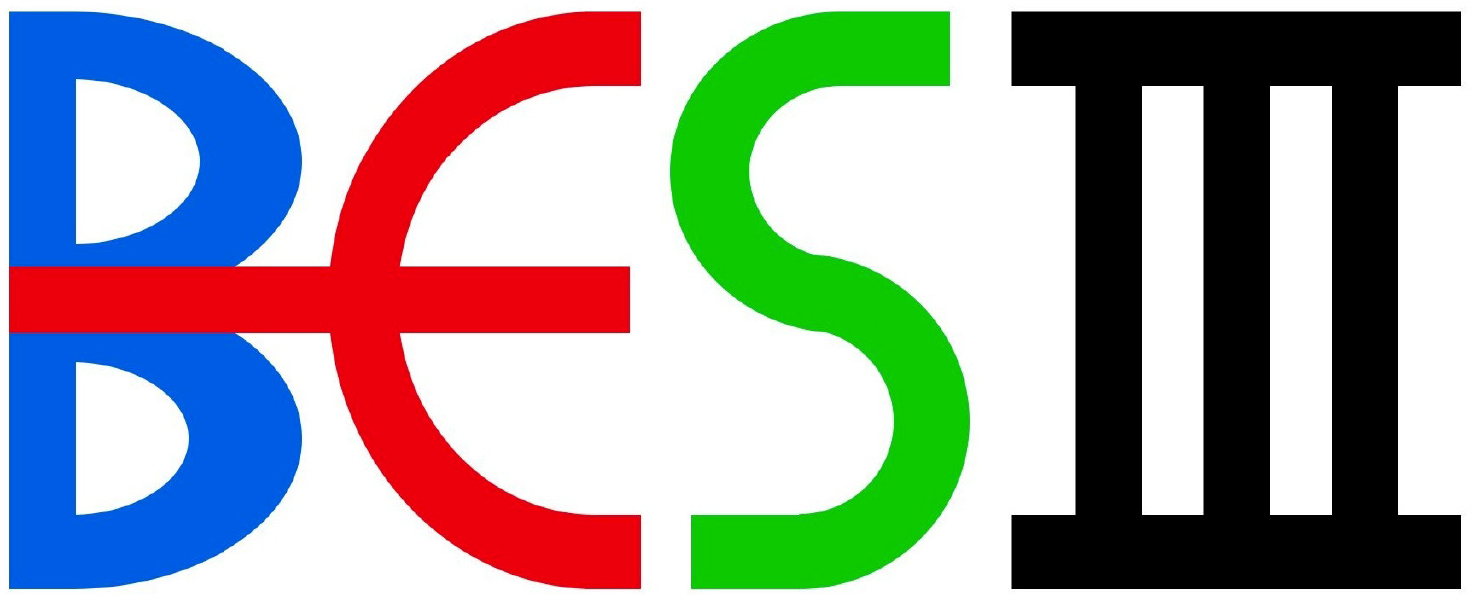}}
\collaboration{The BESIII Collaboration}

\date{\today}
\abstract{The first search for the baryonic decay $ D_{s}^{*+} \to \ p \bar{n} $ is performed using $e^+e^-$ collision data taken at center-of-mass energies between 4.128 and 4.226 GeV,
collected by the BESIII experiment and corresponding to an integrated luminosity of 7.33 fb$^{-1}$. 
No significant signal is observed, and an upper limit on the branching fraction is set to be $1.3\times 10^{-4}$ at the $90\%$ confidence level. 
}

\arxivnumber{}
\begin{document}
\maketitle
\flushbottom

\section{Introduction}
The baryonic decays $ D_{s}^{(*)+} \to \ p \bar{n} $ are weak decays that exchange a $W$ boson,
which are strongly suppressed in the Standard Model (SM) but are very sensitive to new physics beyond the SM.
Intense theoretical and experimental investigations of the ground-state pseudoscalar meson $D_s^+$ 
have been carried out to test the SM.
Under the vacuum-insertion approximation, 
the branching fraction (BF) of $ D_{s}^{+} \to \ p \bar{n} $ at the short-distance level
is predicted to be very small (of the order $10^{-6}$) owing to chiral suppression by the factor $(\frac{m_{\pi}}{m_{D_s}})^2$~\cite{Pham:1980dc, CHEN2008326}, where ${m_{\pi}}$ and ${m_{D_s}}$  are the masses of the pion and $D_s$ meson, respectively. This is a consequence of partial conservation of the axial-vector current~\cite{Pham:1980xe}. It receives long-distance contributions through final-state rescattering of the leading tree and color-suppressed amplitudes. However, the decay $ D_{s}^{+} \to p \bar{n} $ is dominated by long-distance contributions, analogous to the mechanism in its mesonic decays, making it observable~\cite{CHEN2008326}. 
In 2008, the CLEO Collaboration measured the BF of $ D_{s}^{+} \to \ p \bar{n} $ to be $(1.30\pm 0.36^{+0.12}_{-0.16})\times 10^{-3}$ by using 325 pb$^{-1}$ of $e^+e^-$ annihilation data collected at a center-of-mass energy of $\sqrt{s} = $ 4.170 GeV~\cite{CLEO:2008aum}. The BESIII Collaboration confirmed its BF and measured
it to be $(1.21\pm 0.10\pm 0.05)\times 10^{-3}$ by using $e^+e^-$ annihilation data, corresponding to an integrated luminosity of 3.19 fb$^{-1}$, collected at  $\sqrt{s} = $ 4.178 GeV~\cite{dspnbar}. The BF of $ D_{s}^{+} \to \ p \bar{n} $ is of the order $10^{-6}$ at the short-distance level, while experimental measurements give a value of the order $10^{-3}$. This distinct difference between the theoretical and experimental studies of $ D_{s}^{+} \to \ p \bar{n} $ provides motivation to measure the decay of the excited state vector meson $D_{s}^{*+}$, 
which has same final states to explore the physical mechanisms behind the difference.

Although there is no theoretical or experimental study of the weak decay $D_{s}^{*+} \to \ p \bar{n} $, some clues can be found in the studies of the $ D_{s}^{*+}$ meson~\cite{CLEO:2008aum, 2024kxe}. The $W \to e^+ \nu_e$ transition is analogous to the quark-level process $W \to u d $ in the SM.
Considering the {\it{u}} and {\it{d}}  quarks coupling to the baryons {\it{p}} and $\bar{n}$ via two gluons with the coupling constant $\alpha_s^2 \approx 0.3^2 \approx 0.1$~\cite{Particle Data Group:2024cfk}, the BF of the $W \to e^+ \nu_e $ transition could be one order of magnitude larger than $W \to \ p \bar{n} $.
According to the BF of the weak decay $ D_{s}^{*+} \to e^+ \nu_e$ which is $(2.1^{+1.2}_{-0.9}) \times 10^{-5}$~\cite{2024kxe},
the BF of $ D_{s}^{*+} \to \ p \bar{n} $ is estimated to be in the range of $10^{-6}$ and $10^{-7}$ at the short-distance level.
 
This article presents the first search for the baryonic decay $ D_{s}^{*+} \to \ p \bar{n} $ using $e^+e^-$ collision data at $\sqrt{s}=$ 4.128 - 4.226 GeV, collected by the BESIII experiment and corresponding to an integrated luminosity of 7.33 fb$^{-1}$~\cite{4226Gev, 733fb}. The study will improve our understanding of weak decays of charmed mesons.
Throughout this article, charge-conjugate modes are implied.

\section{BESIII detector and data sets}
\label{sec:detector_dataset}
The BESIII detector~\cite{BESIII:2009fln} records symmetric $e^+e^-$ collisions provided by the BEPCII storage ring~\cite{Yu:2016cof}, at center-of-mass energies $\sqrt{s}=$ 1.84 - 4.95 GeV, with a peak luminosity of $1.1\times 10^{33}$ cm$^{-2}$s$^{-1}$ achieved at $\sqrt{s}=$ 3.773 GeV.
The cylindrical core of the BESIII detector covers 93\% of the full solid angle and consists of a helium-based multilayer drift chamber~(MDC), 
a plastic scintillator time-of-flight system~(TOF), and a CsI~(Tl) electromagnetic calorimeter~(EMC), 
which are all enclosed in a superconducting solenoidal magnet providing a 1.0~T magnetic field. 
The solenoid is supported by an octagonal flux-return yoke with resistive plate counter muon identification modules interleaved with steel.
The charged-particle momentum resolution at 1~GeV/$c$ is $0.5\%$, and the $dE/dx$ resolution is $6\%$ for electrons from Bhabha scattering. 
The EMC measures photon energies with a resolution of $2.5\%$ ($5\%$) at 1~GeV in the barrel (end cap) region.
The time resolution in the plastic scintillator TOF barrel region is 68~ps, while that in the end cap region is 110~ps. 
The end cap TOF system was upgraded in 2015 using multigap resistive plate chamber technology, providing a time resolution of 60~ps, which benefits 84\% of the data used in this analysis~\cite{Li:2017jpg, Guo:2017sjt, Cao:2020ibk}. 

Inclusive Monte Carlo~(MC) samples, corresponding to 40 times of the data size, are produced with a {\sc geant4}-based~\cite{GEANT4:2002zbu} software  taking into account the geometric description of the BESIII detector and the detector response. The simulated samples are used to determine the detection efficiencies and estimate backgrounds.
The samples include the production of open charm processes, the initial-state radiation production of vector charmonium(-like) states and the continuum processes incorporated in {\sc kkmc}~\cite{Jadach:2000ir, Jadach:1999vf}. 
All particle decays are modeled with {\sc evtgen}~\cite{Lange:2001uf, Ping:2008zz} using BFs either reported by the Particle Data Group (PDG)~\cite{Particle Data Group:2024cfk}, 
when available, or otherwise estimated with {\sc lundcharm}~\cite{Chen:2000tv, Yang:2014vra}. 
Final state radiation (FSR) from charged particles is incorporated using {\sc photos}~\cite{Richter-Was:1992hxq}. 

For the signal MC samples, we generate 4 million events for the decay $ e^{+}e^{-} \to \ D_{s}^{*+}D_{s}^{-}$. The subsequent process $ D_{s}^{*+} \to \ p \bar{n} $ is simulated by phase space model, which assumes a uniform phase space distribution for the decay products, while the accompanying $ D_{s}^{-}$ meson is allowed to decay inclusively~\cite{Jadach:2000ir, Jadach:1999vf}.

\section{Event selection}
The $ D_{s}^{*+}$ mesons are produced mainly through the process $e^+e^- \to D_{s}^{*+}D_{s}^{-} +c.c.$
We investigate the decay $ D_{s}^{*+} \to \ p \bar{n} $ using the double-tag (DT) method, which was pioneered by the MARK-III collaboration~\cite{MARK-III:1985hbd}.
In an event where a $D_{s}^{-}$ meson is fully reconstructed and comes directly from the $e^+e^-$ collision (called a single-tag (ST) $D_{s}^{-}$ meson), the $ D_{s}^{*+}$ meson decaying to $ \ p \bar{n} $ can be searched in the recoiling system.

The ST $D_s^-$ candidates are selected by reconstructing $D^-_s$ mesons in the three decay modes $D_S^- \to$ $K_{S}^{0}K^{-}$, $K^{+}K^{-}\pi^{-}$ and $\pi^{-}\eta^\prime$. These are chosen on the optimization of the figure of merit $\frac{\epsilon} {1.5+\sqrt{B}}$~\cite{Punzi:2020fsv}, where $\epsilon$ is the signal efficiency obtained by the signal MC sample 
and $B$ is the scaled background yield estimated by the inclusive MC sample. Events, where the signal candidate is reconstructed in the presence of an ST $D_s^-$ meson, are denoted as DT events.

The BF of the decay $ D_{s}^{*+} \to \ p \bar{n} $ is determined by
\begin{linenomath*}
\begin{equation}
    	{\cal B}_{\mathrm{sig}}=\frac{N_{\mathrm{sig}}}{\sum_{\alpha,\,i} N^{\mathrm{\alpha,}\,i}_\mathrm{{tag}}\epsilon^{ \alpha,\,i}_\mathrm{{tag,\,sig}}/\epsilon^{\alpha,\,i}_\mathrm{{tag}}},
	\label{eq:Bsig-gen}
\end{equation}
\end{linenomath*}
where $i$ = 1, 2, 3 represents the three tag modes, $\alpha$ represents the  $\sqrt{s}$ point among the eight energies from 4.128 and 4.226 GeV used in this analysis. $N_\mathrm{{sig}}$ denotes the total signal yield from the three tag channels, where the yield for each channel is summed over all sample groups. For the sample group $\alpha$ and tag mode $i$, $N^{\alpha,\,i}_\mathrm{{tag}}$ is the observed ST yield, $\epsilon^{\alpha,\,i}_\mathrm{{tag,\,sig}}$ and $\epsilon^{\alpha,\,i}_\mathrm{{tag}}$ are the DT and ST efficiencies, respectively, which are obtained from the signal and inclusive MC samples. 

\label{ST-selection}

Charged tracks, except for the daughter tracks of $K^0_S$ candidates,
are selected by requiring the track trajectory to approach the interaction point (IP) 
within $\pm10$ cm along the beam direction and within 1 cm in the plane perpendicular to the beam direction.
In addition, the polar angle ($\theta$) between the direction of the charged track and the beam direction must be
within the detector acceptance by requiring $|\!\cos\theta| < 0.93$.
Charged particle identification (PID) is performed by combining d$E$/dx measurement in the MDC
and flight time measurement in the TOF.
Tracks are identified by the PID likelihood $\mathcal{L}_h$ ($h=\pi,\ K$)  for each hadron hypothesis.
A pion (kaon) candidate is required to satisfy the condition $\mathcal{L}_{\pi(K)}>\mathcal{L}_{K(\pi)}$ and $\mathcal{L}_{\pi(K)}>0$.

The $K^0_S$ candidates are reconstructed via $K^0_S\to\pi^+\pi^-$ using charged tracks assumed to be pions without imposing PID criteria,
and the daughter tracks are required to have a trajectory that approaches the IP within $\pm20$ cm
along the beam direction and $|\!\cos\theta| <0.93$.
A vertex-constrained fit is performed to the oppositely charged track pairs of $K^0_S$ candidates, 
and the resulting track parameters are used to calculate the $\pi^+\pi^-$ invariant mass 
which is required to be within [0.487, 0.511] GeV/$c^2$.
To further suppress the combinatorial background, the $\chi^2$ of the vertex fit is required to be less than 100.

Photon candidates are selected from the electromagnetic showers,
which have a minimum energy of 25 MeV in the EMC barrel region ($|\!\cos\theta|< 0.80$) or 50 MeV in the EMC end cap region ($0.86 < |\!\cos\theta| < 0.92$).
To reduce the photon candidates that result from electronic noise or bremsstrahlung, the time of the shower measured by the EMC is required to be less than 700 ns after the beam collision.
The opening angle between a photon and the closest charged track is required to be more than 10$^\circ$,
which is used to remove electrons, hadronic showers, and photons from FSR.

The $\eta^{\prime}$ candidates are reconstructed via the decay mode $\pi^+\pi^-\eta$
by requiring the corresponding invariant mass to be within [0.946, 0.970] GeV/$c^2$.
The $\eta$ candidates are reconstructed via $\eta\to\gamma\gamma$, 
with each photon having an energy greater than 100 MeV, and the photon pairs required to 
have an invariant mass with [0.500, 0.570] GeV/$c^2$.
To improve the momentum resolution, a kinematic fit is performed by 
constraining the $\gamma\gamma$ invariant mass to the nominal value of the $\eta$ meson from the PDG~\cite{Particle Data Group:2024cfk}, 
and the $\chi^2$ of the fit is required to be less than 30 to reject the combinatorial background.

\begin{table}[t!]
	\begin{center}
	\caption{Requirements on $M_{D_s^-}$ for various tag modes.}
	\label{ds_mass_range}
		\begin{tabular}{lc}\hline
		 Tag mode  & Mass window (GeV/$c^{2}$)      \\
		\hline 
		 $\dstoksk$	     &[1.948, 1.991]	\\
		$\dstokkpi$	      &[1.950, 1.986]	\\
		$\dstopietapgam$  &[1.938, 1.997]\\
			\hline
		\end{tabular}
	\end{center}
\end{table}

\begin{figure}[htbp]
\centering
\includegraphics[width=0.95\textwidth]
{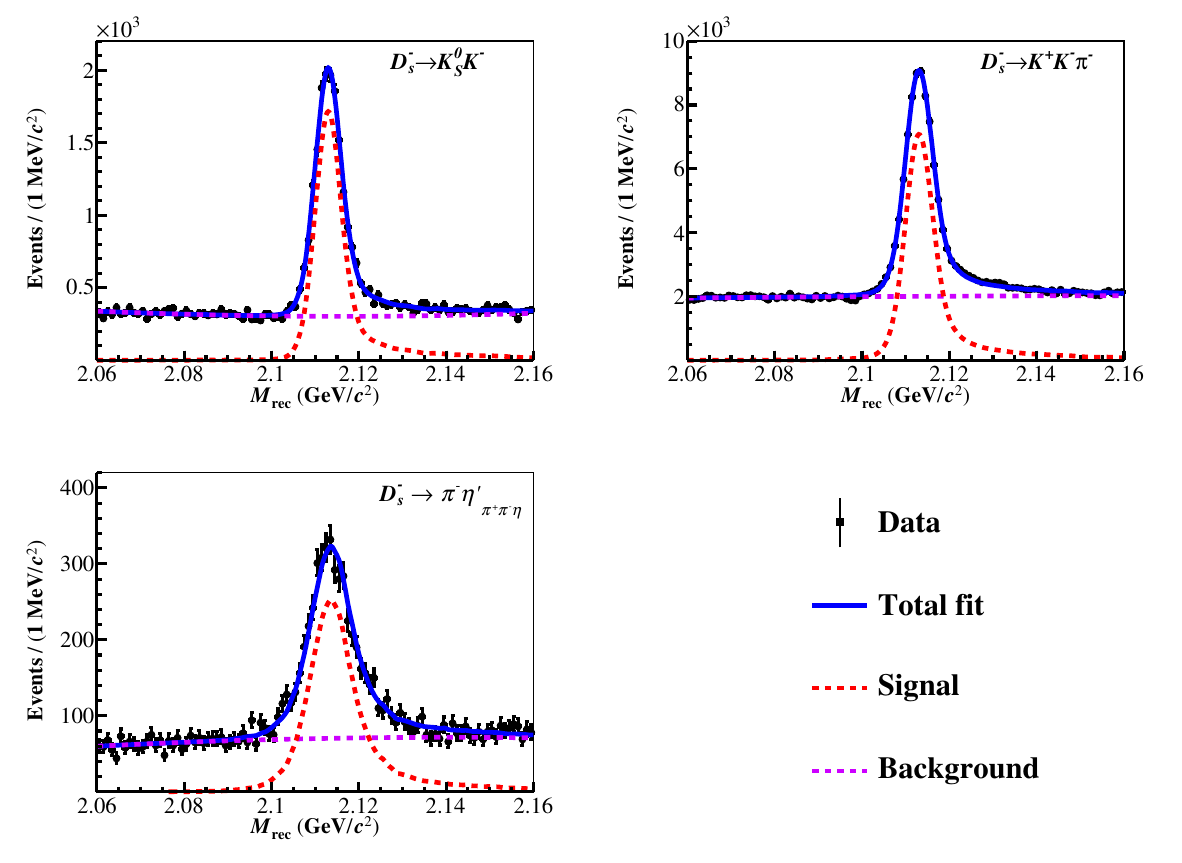}

\caption{The distributions of $M_{\rm rec}$ from data sample at $\sqrt{s}=4.178$ GeV for the three tag modes. The dots with error bars are from data. The solid blue line is the fit result. The dashed red line is the signal shape, and the dashed purple line is the background shape. }
\label{fit:Mass-data-Ds_4180}
\end{figure}

If there are multiple combinations for ST candidates, we choose the one with $M_{D_s^-}$ closest to the known $D_s^-$ mass from the PDG~\cite{Particle Data Group:2024cfk} as the best candidate.
Non-$D_{s}^{*+}D_{s}^{-}$ events are further suppressed by requiring that $M_{D_s^-}$ agrees with the known $D_s^-$ mass within $3 \sigma$ of the $M_{D_s^-}$ distribution of each tag mode as given in Table~\ref{ds_mass_range}.
The recoiling mass $M_{\rm rec}$ of the ST $D_s^-$ is  defined as
\begin{linenomath*}
\begin{equation}
\begin{aligned}
	\begin{array}{lr}
	M_{\rm rec} = \sqrt{\left(E_{\rm cm}/c^2 - \sqrt{|\vec{p}_{D_{s}^-}/c|^{2}+m_{D_{s}^-}^{2}}\right)^{2} - |\vec{p}_{D_{s}^-}/c | ^{2}} \; , \label{eq:mrec}
		\end{array}
\end{aligned}\end{equation}
\end{linenomath*}
where $E_{\rm cm}$ is the  $\sqrt{s}$ of the data sample, $\vec{p}_{D_{s}^-}$ is the three-momentum of the $D_{s}^{-}$ candidate in the $e^+e^-$ center-of-mass frame, and $m_{D_{s}^-}$ is the known  $D_{s}^{-}$ mass.
The $M_{\rm rec}$ distribution for the signal process $e^+e^- \to D_{s}^{*+}D_{s}^{-}$ will form a peak around the $D_s^{*+}$ mass~\cite{Particle Data Group:2024cfk}, while the background processes (e.g. $e^+e^- \to \gamma D_{s}^{+}D_{s}^{-}$, $e^+e^- \to \pi^{0} D_{s}^{+}D_{s}^{-}$) produce flat distributions. The ST yield for each tag mode is determined by fitting the corresponding $M_{\rm rec}$ spectrum. In these fits, the signal is modeled using an MC simulated shape convolved with a Gaussian function which accounts for resolution differences between data and MC simulation. The parameters of these Gaussian functions are treated as free parameters in the fit.
The background is described by a second-order Chebyshev polynomial.
The yields for the three ST modes are listed in Table~\ref{ST-yield}. As an example, the fits to the ST candidates from the data sample at $\sqrt{s} = 4.178$~GeV are shown in Fig.~\ref{fit:Mass-data-Ds_4180}. The same ST selection criteria and fitting approach applied to data are used to analyze the inclusive MC samples to extract the ST efficiency as shown in Table~\ref{ST-eff}.
\renewcommand{\figurename}{FIG.}

\begin{table}[htp]
	\begin{center}
		\caption{The ST yields ($N^{\mathrm{\alpha},\,i}_\mathrm{{tag}}$) obtained from  data at $\sqrt{s}=$ 4.128 - 4.226 GeV. The uncertainties are statistical.}
		\label{ST-yield}
         \vspace{2pt} 
		
			\begin{tabular}{l cccc}
   \hline
				Tag mode & $N^{\rm 4128}_{\rm tag}$ & $N^{\rm 4157}_{\rm tag}$ &$N^{\rm 4178}_{\rm tag}$ & $N^{\rm 4189}_{\rm tag}$  \\\hline
				$\dstoksk$	    &$1224\pm44$ &$1989\pm63$ &$15644\pm179$&$2765\pm 77$ \\
				$\dstokkpi$	    &$5198\pm112$&$8493\pm151$&$69838\pm424$&$12069\pm182$ \\
				$\dstopietapgam$&$281\pm28$  &$476\pm41$  &$3725 \pm103$&$681\pm46$\\\hline
				Tag mode & $N^{\rm 4199}_{\rm tag}$ & $N^{\rm 4209}_{\rm tag}$ & $N^{\rm 4219}_{\rm tag}$ & $N^{\rm 4226}_{\rm tag}$\\\hline
				$\dstoksk$	    &$2529\pm76$&$2600\pm81$&$2212\pm77$&$3247\pm95$\\
				$\dstokkpi$     &$11049\pm181$&$11401\pm186$&$8709\pm169$&$14310\pm240$\\
				$\dstopietapgam$&$715\pm51$&$708\pm58$&$541\pm45$&$796\pm52$\\\hline
			\end{tabular}
	\end{center}
\end{table}

\begin{table}[htp]
	\begin{center}
		\caption{The ST efficiencies ($\epsilon^{\alpha,\,i}_\mathrm{{tag}}$) obtained from inclusive MC samples at $\sqrt{s}=$ 4.128 - 4.226 GeV. The uncertainties are statistical.}
		\label{ST-eff}
        \vspace{2pt} 
			\begin{tabular}{l ccccc}\hline
				Tag mode & $\epsilon^{\rm 4128}_{\rm tag}$(\%) & $\epsilon^{\rm 4157}_{\rm tag}$(\%)& $\epsilon^{\rm 4178}_{\rm tag}$(\%)& $\epsilon^{\rm 4189}_{\rm tag}$(\%)
				\\\hline
$\dstoksk$	    &$46.26\pm0.28$&$48.37\pm0.24$&$48.12\pm0.08$&$48.03\pm0.21$\\
$\dstokkpi$	    &$38.69\pm0.13$&$40.87\pm0.11$&$41.10\pm0.04$&$40.47\pm0.09$\\
$\dstopietapgam$&$18.52\pm0.27$&$19.77\pm0.23$&$19.96\pm0.08$&$20.24\pm0.20$\\\hline
Tag mode& $\epsilon^{\rm 4199}_{\rm tag}$(\%)& $\epsilon^{\rm 4209}_{\rm tag}$(\%)&$\epsilon^{\rm 4219}_{\rm tag}$(\%)& $\epsilon^{\rm 4226}_{\rm tag}$(\%)\\\hline
$\dstoksk$	    &$49.83\pm0.22$&$49.55\pm0.22$&$49.28\pm0.31$&$48.09\pm0.20$\\
$\dstokkpi$	    &$42.00\pm0.10$&$41.82\pm0.10$&$41.58\pm0.11$&$40.44\pm0.09$\\
$\dstopietapgam$&$20.37\pm0.21$&$20.23\pm0.22$&$20.52\pm0.28$&$19.96\pm0.20$\\\hline
			\end{tabular}
	\end{center}
\end{table}

On the recoiling side of the ST $D_s^-$ meson, we require that there is only one residual charged track to be identified as a proton by the PID likelihoods which must satisfy $\mathcal{L}(p)>\mathcal{L}(K)$, $\mathcal{L}(p)>\mathcal{L}(\pi)$ and $\mathcal{L}(p) \geq 0.001$.

The signal is separated from combinatorial backgrounds using the square of the missing mass defined as
\begin{linenomath*}
    \begin{equation}
	\label{recoilingmass}
	M_{\rm miss}  = \sqrt{\frac{\left| E_{\rm cm} - \sum_{k} E_{\it k}  \right|^{2}} {c^{4}} -  \frac {|- \sum_{k} \vec{p}_{k}|^{2}} {c^{2}}},
    \end{equation}
\end{linenomath*}
where $E_{\rm cm}$ is the center-of-mass  energy, $E_{\it k}$ and $\vec{p}_{k}$ are the energy and momentum of the particle {\it{k}}, with {\it{k}} runing over the particles from the $ D_{s}^-$ decays and proton of the signal side.

To improve the resolution, we perform kinematic fits to the signal candidates. The four-momenta of the tagged $D_{s}$ and the $D_{s}^{*}$, as well as the total system four-momentum, are constrained in the fit, whereas the four-momentum of the missing neutron is allowed to vary.

The potential backgrounds are classified into (a) non-$ D_{s}^-$ background and (b) real-$ D_{s}^-$ background. The background (a) is dominated by continuum processes with a proton and an antineutron in the final state. The background (a) contains genuine neutrons and therefore appear as a signal-like peak around the neutron mass in the $M_{\rm miss}$ spectrum centered. However, due to the incorrectly tagged $ D_{s}^-$, its distribution in $M_{\rm rec}$ remains flat. Finally, we perform a two-dimensional (2D) fit to the $M_{\rm rec}$ versus $M_{\rm miss}$ distributions to extract the signal yield. The background (b) mostly comes from the process $ D_{s}^{+} \to \ p \bar{n} $.
The $ D_{s}^{*+} \to \ p \bar{n} $ and $ D_{s}^{+} \to \ p \bar{n} $ processes exhibit distinct peak positions in the $M_{\rm miss}$ spectrum.
The decay $ D^{*+}_{s} \to \ p \bar{n} $ forms a peak at the neutron mass in the $M_{\rm miss}$ distribution. In contrast, the $M_{\rm miss}$ distribution for $ D_{s}^{+} \to \ p \bar{n} $ is concentrated in the region above 1 GeV/$c^{2}$. To suppress this background we choose $M_{\rm miss} < 1.00 $  GeV/$c^{2}$ as the fit range. The signal retention rate of this $M_{\rm miss}$ requirement is $97\%$.
The properties of the backgrounds are validated by studying the inclusive MC samples.

\section{Results}
\label{BFSelection} 

\begin{figure}[htbp]
	\centering
   {\includegraphics[width=0.46\textwidth]{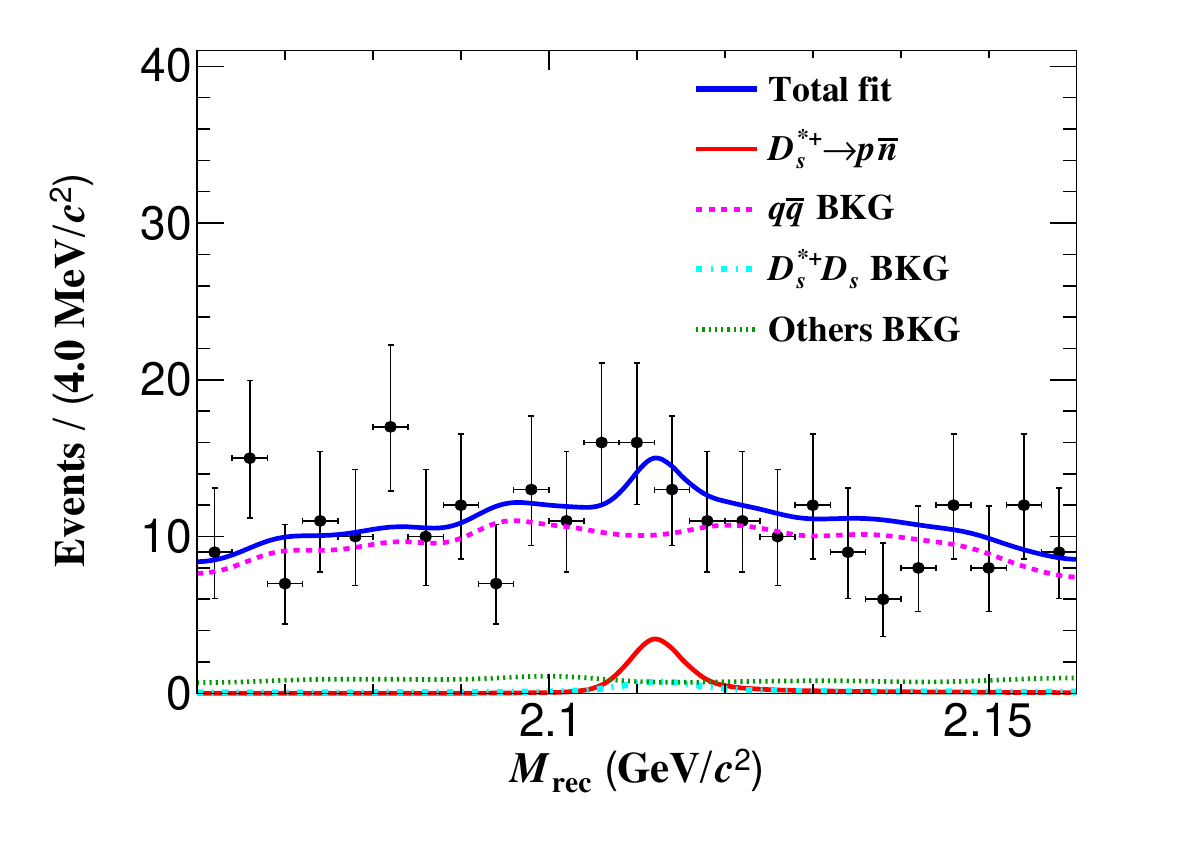}}
	{\includegraphics[width=0.46\textwidth]{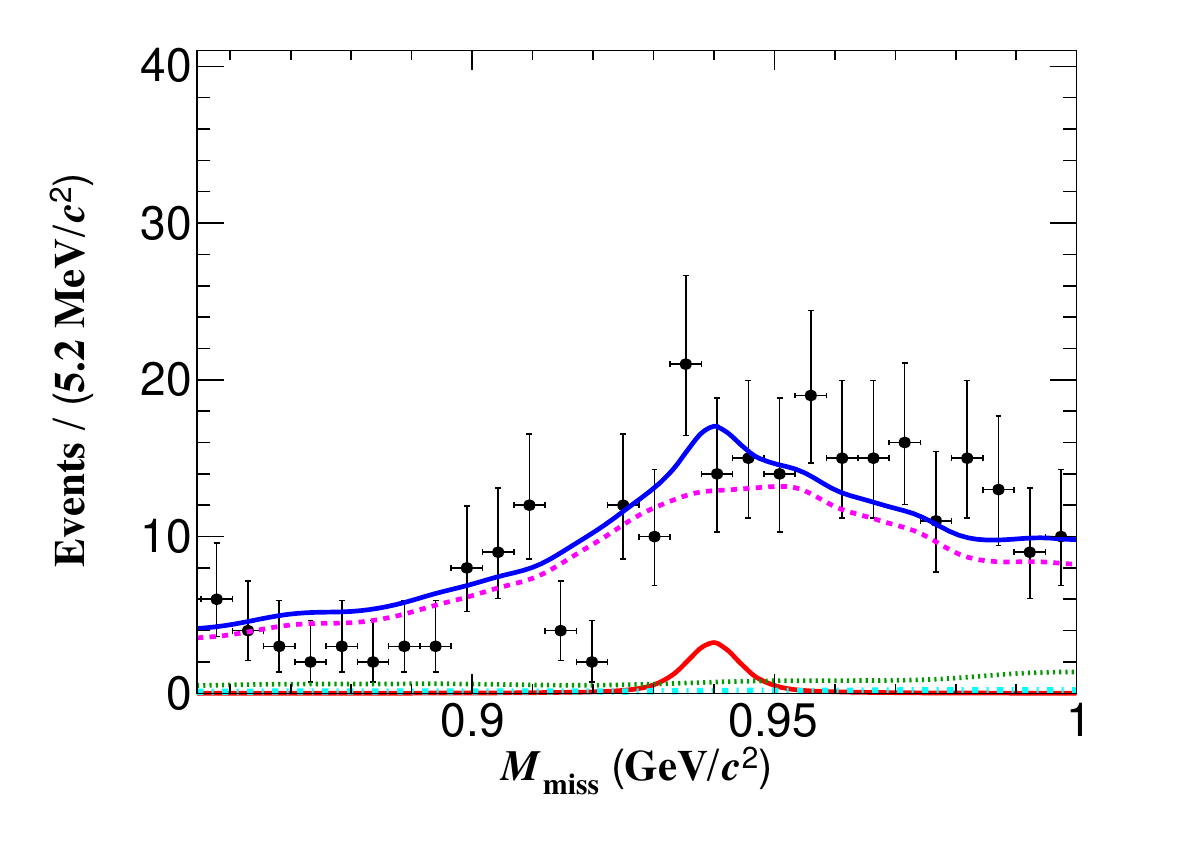}}
	\caption{Projections on $M_{\rm rec}$ and $M_{\rm miss}$ of the 2D fit result from data sample at $\sqrt{s}=$ 4.128 - 4.226 GeV. The dots with error bars represent the data. The solid blue line represents the total fit. The dashed magenta line represents the background of  $ q\bar{q} $. The solid red line represents the signal shape. The dash-dotted blue line represents the background of $ D_{s}^{*+} D_{s}$. The dotted green line represents the residual background not covered by other categories.}
	\label{data_twofit}
\end{figure}

After imposing all above selection criteria, 
a 2D unbinned maximum likelihood fit on the $M_{\rm rec}$ versus $M_{\rm miss}$ distributions of selected $ D_{s}^{*+} \to \ p \bar{n} $ candidates is shown in Fig.~\ref{data_twofit}. 
The signal shape is described by an MC-simulated 2D probability density function (PDF) convolved with a  Gaussian function, and the shape of the background components are derived from the inclusive MC samples.

From the 2D fit, the number of signal events is determined to be $8.5\pm5.6$, 
with a statistical significance of 1.9$\sigma$. The fit quality is evaluated to be $\chi^{2}/\text{ndf} =29.7/25 = 1.19$, where $\text{ndf}$ is the number of degree of freedom. The statistical significance of the $ D_{s}^{*+} \to \ p \bar{n} $ decay is evaluated using $\sqrt{-2{\rm ln}(\mathcal {L}_0/\mathcal {L}_{\rm max})}$, where $\mathcal {L}_{\rm max}$ is the maximum likelihood of the nominal fit and $\mathcal {L}_0$ is the likelihood of the fit without the signal component.
The corresponding DT efficiencies for each tag mode are determined and listed in Table~\ref{DT-eff}. 

\begin{table}[htp]
	\begin{center}
		\caption{The DT efficiencies ($\epsilon^{ \alpha,\,i}_\mathrm{{tag,\,sig}}$) obtained from signal MC samples at $\sqrt{s}=$ 4.128 - 4.226 GeV. The uncertainties are statistical.}
		\label{DT-eff}
        \vspace{2pt} 
			\begin{tabular}{l ccccc}\hline
				Tag mode & $\epsilon^{\rm 4128}_\mathrm{{tag,\,sig}}$(\%) & $\epsilon^{\rm 4157}_\mathrm{{tag,\,sig}}${(\%)}& $\epsilon^{\rm 4178}_\mathrm{{tag,\,sig}}${(\%)}& $\epsilon^{\rm 4189}_\mathrm{{tag,\,sig}}${(\%)}
				\\\hline
$\dstoksk$	    &$43.64\pm0.24$&$43.60\pm0.24$&$42.99\pm0.24$&$42.90\pm0.25$\\
$\dstokkpi$	    &$38.09\pm0.10$&$37.33\pm0.10$&$36.48\pm0.10$&$36.22\pm0.10$\\
$\dstopietapgam$&$18.27\pm0.24$&$18.37\pm0.24$&$17.80\pm0.23$&$17.22\pm0.22$\\\hline
Tag mode& $\epsilon^{\rm 4199}_\mathrm{{tag,\,sig}}$(\%)& $\epsilon^{\rm 4209}_\mathrm{{tag,\,sig}}${(\%)}&$\epsilon^{\rm 4219}_\mathrm{{tag,\,sig}}${(\%)}& $\epsilon^{\rm 4226}_\mathrm{{tag,\,sig}}${(\%)}\\\hline
$\dstoksk$	    &$42.24\pm0.25$&$40.52\pm0.25$&$38.10\pm0.24$&$38.50\pm0.24$\\
$\dstokkpi$	    &$35.83\pm0.10$&$34.22\pm0.10$&$32.85\pm0.10$&$32.76\pm0.10$\\
$\dstopietapgam$&$17.61\pm0.23$&$16.50\pm0.23$&$16.23\pm0.22$&$16.52\pm0.22$\\\hline
				
			\end{tabular}
		
	\end{center}
\end{table}

\begin{figure}[t!]
          \centering
           {\includegraphics[width=7.0cm]{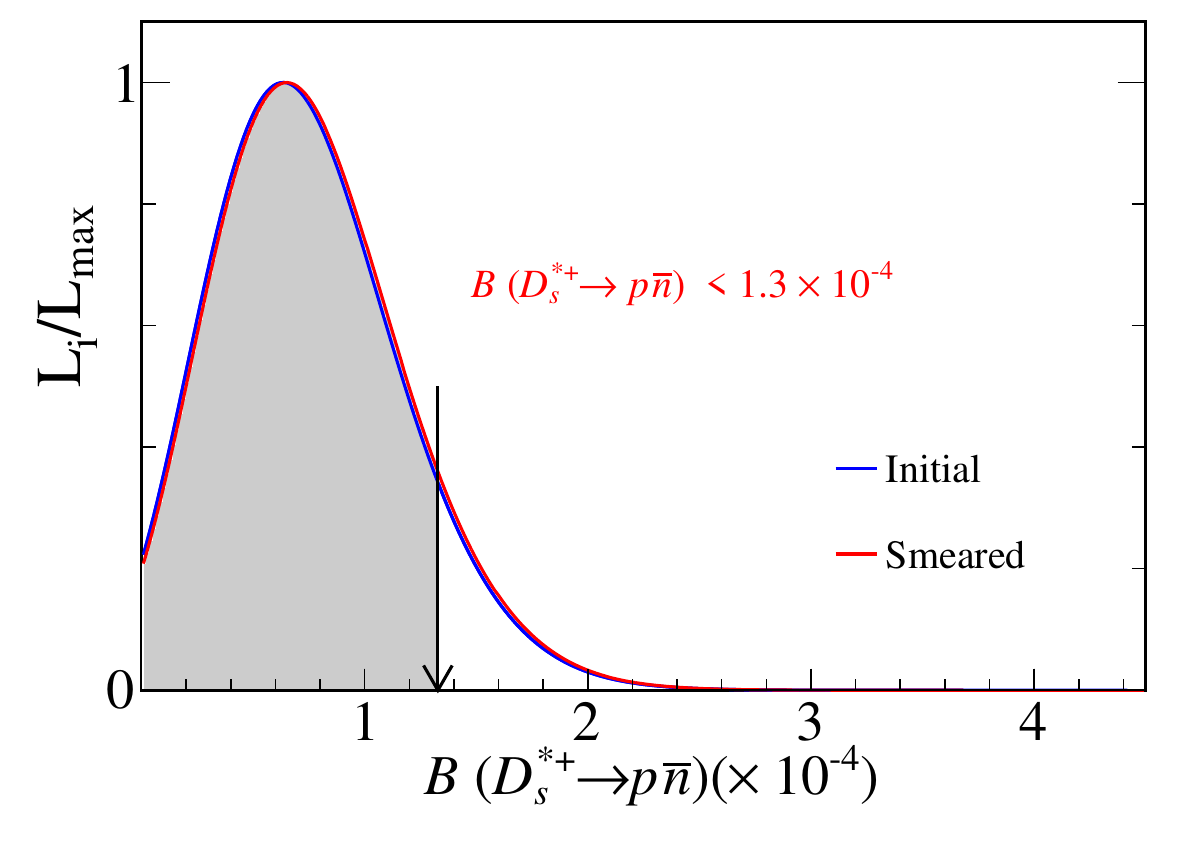}}
		\caption{Normalized likelihood versus the BF of $ D_{s}^{*+} \to \ p \bar{n} $. 
		The blue line is the likelihood curve for the nominal fit model, while the red line represents the likelihood curve that gives the upper limit after incorporating the systematic uncertainty. 
		  The grey shaded area indicates the $90\%$ confidence level of the BF of $ D_{s}^{*+} \to \ p \bar{n} $. 
		}
    \label{fig:bf2}
\end{figure}

Since no significant signal of the decay $ D_{s}^{*+} \to \ p \bar{n} $ is found, 
an upper limit on the signal yields is determined after incorporating both additive and multiplicative systematic uncertainties via a likelihood scan method, following Ref.~\cite{BESIII:2022jcm}. Additive uncertainties are studied by repeating the fit with alternative models and choosing the most conservative upper limit. Then the multiplicative systematic uncertainty $\sigma_{\epsilon}$ is incorporated by following Refs.~\cite{Stenson:2006gwf, Liu:2015uha}.
\begin{linenomath*}
    \begin{equation}
	L(n)\propto\int_{0}^{1}L(n\cdot\frac{\epsilon}{\epsilon_{0}})e^{-\frac{(\epsilon-\epsilon_{0})^2}{2(\sigma _{\epsilon}\epsilon_{0})^2}}d\epsilon,
    \end{equation}
\end{linenomath*}
where $\epsilon_{0}$ is the nominal efficiency based on the signal MC sample.
The likelihood distributions with and without incorporating the systematic uncertainties for the upper limit on the BF of $ D_{s}^{*+} \to \ p \bar{n} $ are shown in Fig.~\ref{fig:bf2}. The upper limit, including all systematic uncertainties, is found to be $1.3\times 10^{-4}$ at the 90\% confidence level.

\section{Systematic uncertainties}
The systematic uncertainties in the BF measurement are discussed below.
The additive systematic uncertainties only affect the fitted signal yield, while multiplicative systematic uncertainties affect the detection efficiency. 
Most systematic uncertainties related to the determination of the efficiency for reconstructing the ST side cancel out in the BF measurement due to the DT technique. 

The additive systematic uncertainties are discussed below. Among the results of these fits, the one corresponding to the largest upper limit on the BF is chosen. 
\begin{itemize}

\item Signal shape. The systematic uncertainty related with the signal shape, is studied by the fit without the Gaussian function convolved.

\item Background shape. For background shape, we alter the MC-simulated shapes by varying the relative fractions of the two major background components from $q\bar q$ and $D_s^{*+}D_s^-$ by $\pm 30\%$ and $\pm 12\%$~\cite{BESIII:2021xox}, which are the statistical uncertainties of the individual cross sections in data. 
\end{itemize}

The multiplicative systematic uncertainties of relative values are discussed below.
\begin{itemize}
\item ST yield. 
The total ST yield of all ST modes is $181200\pm706$,
		resulting in the statistical uncertainty of 0.3\%.
Here, we only consider the statistical fluctuation related to the background of the ST side,
as it is not correlated with the DT sample directly. 
Hence, we take the systematic uncertainty to be $0.3\%$.
\item Tracking and PID. 
The efficiencies for proton tracking and PID have been studied based on $e^+e^- \to \pi^{+}  \pi^{-} p \overline{p} $ in Ref.~\cite{dspnbar}.~The data-MC differences from the control sample, $0.5\%$ for proton tracking and $1.5\%$ for proton PID, are taken as the systematic uncertainties.

\item MC statistics. The uncertainty due to the limited signal MC sample size is evaluated by $\frac{1}{\sqrt{N}} \sqrt{\frac{1-\epsilon}{\epsilon}}$, 
where $\epsilon$ is the detection efficiency, {\it{N}}  ~is the total number of the generated MC events. 
This uncertainty is found to be $0.1\%$.
\end{itemize}

The systematic uncertainties are summarised in Table~\ref{abs:bfsys}.
Summing the contributions discussed above in quadrature leads to a total multiplicative systematic uncertainty of $1.6\%$.

\begin{table}[t!]
  \centering
  	\caption{Systematic uncertainties in the BF measurement of $ D_{s}^{*+} \to \ p \bar{n} $.}
  \begin{tabular}{l|l|c}
  
    \hline
    Category &Source   &$\text{BF}^{\text{UL}} ~(90\%\ \text{C.L.}) ~\times 10^{-4}$ \\
   \hline
    \multirow{2}{*}{Additive} 
    &Signal shape  &1.3\\
    &Background shape &1.3\\
\hline
    Category &Source   &Uncertainty (\%) \\
  \hline
    \multirow{6}{*}{Multiplicative}&ST yield      &0.3\\
    &Tracking                                &0.5\\
    &PID                                     &1.5\\
  
    &MC statistics                           &0.1\\
    \cline{2-3}
    &Total                                   &1.6\\
    
    \cline{2-3}

    \hline
  \end{tabular}
  \label{abs:bfsys}
\end{table}

\section{Conclusion}

This article presents the first search for the decay $D_{s}^{*+} \to\ p \bar{n}$ using $e^+e^-$ collision data at $\sqrt{s}=$ 4.128 - 4.226 GeV, collected by the BESIII experiment and corresponding to an integrated luminosity of 7.33 fb$^{-1}$. Employing the double-tag method by reconstructing single-tag $D_s^{-}$ mesons in three hadronic decay modes: $D_s^{-} \to K_S^0 K^{-}$, $D_s^{-} \to K^{-} K^{+} \pi^{-}$ and $\dstopietapgam$, a 2D unbinned maximum likelihood fit on the $M_{\text{rec}}$ and $M_{\text{miss}}$ distributions is performed.
An upper limit on the BF of the decay $ D_{s}^{*+} \to \ p \bar{n} $ is determined to be $1.3\times 10^{-4}$ at the 90\% confidence level. There is no theoretical calculation of the BF to compare this to at this time.

\acknowledgments

The BESIII Collaboration thanks the staff of BEPCII (https://cstr.cn/31109.02.BEPC) and the IHEP computing center for their strong support. This work is supported in part by National Key R\&D Program of China under Contracts Nos. 2023YFA1606000, 2023YFA1606704, 2025YFA1613900; National Natural Science Foundation of China (NSFC) under Contracts Nos. 12305102, 12275067, 12275068, 11635010, 11935015, 11935016, 11935018, 12025502, 12035009, 12035013, 12061131003, 12192260, 12192261, 12192262, 12192263, 12192264, 12192265, 12221005, 12225509, 12235017, 12342502, 12361141819, 12535005; Outstanding Youth Foundation of Henan Province No. 252300421219; Science and Technology R$\&$D Program Joint Fund Project of Henan Province No.225200810030, Science and Technology Innovation Leading Talent Support Program of Henan Province; 
Natural Science Foundation of Henan Province Nos. 252300421491, 242300420250; Natural Science Foundation of Hunan Province under Contract No.~2026JJ20021; Hunan Province Major Basic Research Projects under Contract No.~2026JC0006; Open Funding(QLPL2026P01) under Key Laboratory of Quark and Lepton Physics (MOE), Central China Normal University, Wuhan 430079, China; Innovative Platform for Quark Matter Science and Advanced Detection Technology funded under Ministry of Education's "Two-New" Initiative; the Chinese Academy of Sciences (CAS) Large-Scale Scientific Facility Program; the Strategic Priority Research Program of Chinese Academy of Sciences under Contract No. XDA0480600; CAS under Contract No. YSBR-101; 100 Talents Program of CAS; The Institute of Nuclear and Particle Physics (INPAC) and Shanghai Key Laboratory for Particle Physics and Cosmology; Agencia Nacional de Investigación y Desarrollo de Chile (ANID), Chile under Contract No. ANID CCTVal CIA250027; ERC under Contract No. 758462; German Research Foundation DFG under Contract No. FOR5327; Istituto Nazionale di Fisica Nucleare, Italy; Knut and Alice Wallenberg Foundation under Contracts Nos. 2021.0174, 2021.0299, 2023.0315; Ministry of Development of Turkey under Contract No. DPT2006K-120470; National Research Foundation of Korea under Contract No. NRF-2022R1A2C1092335; National Science and Technology fund of Mongolia; Polish National Science Centre under Contract No. 2024/53/B/ST2/00975; STFC (United Kingdom); Swedish Research Council under Contract No. 2019.04595; U. S. Department of Energy under Contract No. DE-FG02-05ER41374.

\clearpage

M.~Ablikim$^{1}$\BESIIIorcid{0000-0002-3935-619X},
M.~N.~Achasov$^{4,c}$\BESIIIorcid{0000-0002-9400-8622},
P.~Adlarson$^{83}$\BESIIIorcid{0000-0001-6280-3851},
X.~C.~Ai$^{89}$\BESIIIorcid{0000-0003-3856-2415},
C.~S.~Akondi$^{31A,31B}$\BESIIIorcid{0000-0001-6303-5217},
R.~Aliberti$^{39}$\BESIIIorcid{0000-0003-3500-4012},
A.~Amoroso$^{82A,82C}$\BESIIIorcid{0000-0002-3095-8610},
Q.~An$^{78,65,\dagger}$,
Y.~H.~An$^{89}$\BESIIIorcid{0009-0008-3419-0849},
M.~S.~Anderson$^{39}$\BESIIIorcid{0009-0008-1550-2632}
% ...
\\
\vspace{0.2cm}
(BESIII Collaboration)\\
\vspace{0.2cm} {\it
$^{1}$ Institute of High Energy Physics, Beijing 100049, People's Republic of China\\
% ...
$^{4}$ Budker Institute of Nuclear Physics SB RAS (BINP), Novosibirsk 630090, Russia\\
% ...
$^{31}$ INFN Sezione di Ferrara, (A)INFN Sezione di Ferrara, I-44122, Ferrara, Italy; (B)University of Ferrara, I-44122, Ferrara, Italy\\
% ...
$^{39}$ Johannes Gutenberg University of Mainz, Johann-Joachim-Becher-Weg 45, D-55099 Mainz, Germany\\
% ...
$^{65}$ State Key Laboratory of Particle Detection and Electronics, Beijing 100049, Hefei 230026, People's Republic of China\\
% ...
$^{78}$ University of Science and Technology of China, Hefei 230026, People's Republic of China\\
% ...
$^{82}$ University of Turin and INFN, (A)University of Turin, I-10125, Turin, Italy; (B)University of Eastern Piedmont, I-15121, Alessandria, Italy; (C)INFN, I-10125, Turin, Italy\\
$^{83}$ Uppsala University, Box 516, SE-75120 Uppsala, Sweden\\
% ...
$^{89}$ Zhengzhou University, Zhengzhou 450001, People's Republic of China\\

\vspace{0.2cm}
$^{\dagger}$ Deceased\\
% ...
$^{c}$ Also at the Novosibirsk State University, Novosibirsk, 630090, Russia\\
% ...
$^{n}$ Also at Helmholtz Institute Mainz, Staudinger Weg 18, D-55099 Mainz, Germany\\
% ...

}

\end{document}